\documentclass[journal=nalefd]{achemso}
\pdfoutput=1
\usepackage{amssymb}

\author{N. L. B. Ziino}
\affiliation[Center for Quantum Devices]{Center for Quantum Devices, Niels Bohr Institute, University of Copenhagen, Universitetsparken 5, DK-2100 Copenhagen, Denmark}

\author{P. Krogstrup}
\affiliation[Center for Quantum Devices]{Center for Quantum Devices, Niels Bohr Institute, University of Copenhagen, Universitetsparken 5, DK-2100 Copenhagen, Denmark}
\author{M. H. Madsen}

\author{E. Johnson}
\affiliation[Center for Quantum Devices]{Center for Quantum Devices, Niels Bohr Institute, University of Copenhagen, Universitetsparken 5, DK-2100 Copenhagen, Denmark}
\alsoaffiliation[DTU Wind Energy]{DTU Wind Energy, DTU Campus Ris{\o}, Technical University of Denmark}

\author{J. B. Wagner}
\affiliation[Center for Electron Nanoscopy]{Center for Electron Nanoscopy, Technical University of Denmark}

\author{C. M. Marcus}
\affiliation[Center for Quantum Devices]{Center for Quantum Devices, Niels Bohr Institute, University of Copenhagen, Universitetsparken 5, DK-2100 Copenhagen, Denmark}

\author{J. Nyg{\aa}rd}
\affiliation[Center for Quantum Devices]{Center for Quantum Devices, Niels Bohr Institute, University of Copenhagen, Universitetsparken 5, DK-2100 Copenhagen, Denmark}

\author{T. S. Jespersen}
\affiliation[Center for Quantum Devices]{Center for Quantum Devices, Niels Bohr Institute, University of Copenhagen, Universitetsparken 5, DK-2100 Copenhagen, Denmark}
\email{tsand@fys.ku.dk}

\title{Epitaxial aluminum contacts to InAs nanowires}

\begin{document}


%
%
%
%

\begin{abstract}
We report a method for making epitaxial superconducting contacts to semiconducting nanowires. The temperature and gate characteristics demonstrate barrier-free electrical contact, and the properties in the superconducting state are investigated at low temperature.  Half-covering aluminum contacts are realized without the need of lithography and we demonstrate how to controllably insert high-band gap layers in the interface region. These developments are relevant to hybrid superconductor-nanowire devices that support Majorana zero energy states.
\end{abstract}

Semiconducting indium arsenide (InAs) nanowires (NWs) have been implemented as the active elements in nanoscale electrical devices for a wide range of studies. Their high electron mobilities and low effective mass provide a good basis for superior field effect transistors \cite{Samuelson:2004}, their surface transport channels and high surface-to-volume ratio make them noteworthy candidates for use in chemical sensors\cite{Du:2009}, and their strong spin-orbit coupling \cite{Hansen:2005,Fasth:2007}, large $g$-factors\cite{Bjork:2005} and relative ease of contacting to superconducting (SC) contacts has led to a number of breakthroughs in quantum transport \cite{NadjPerge:2010, Doh:2005, Vandam:2006, Hofstetter:2009}.\\ 
\begin{figure}
        \centering
        \includegraphics[width=8cm]{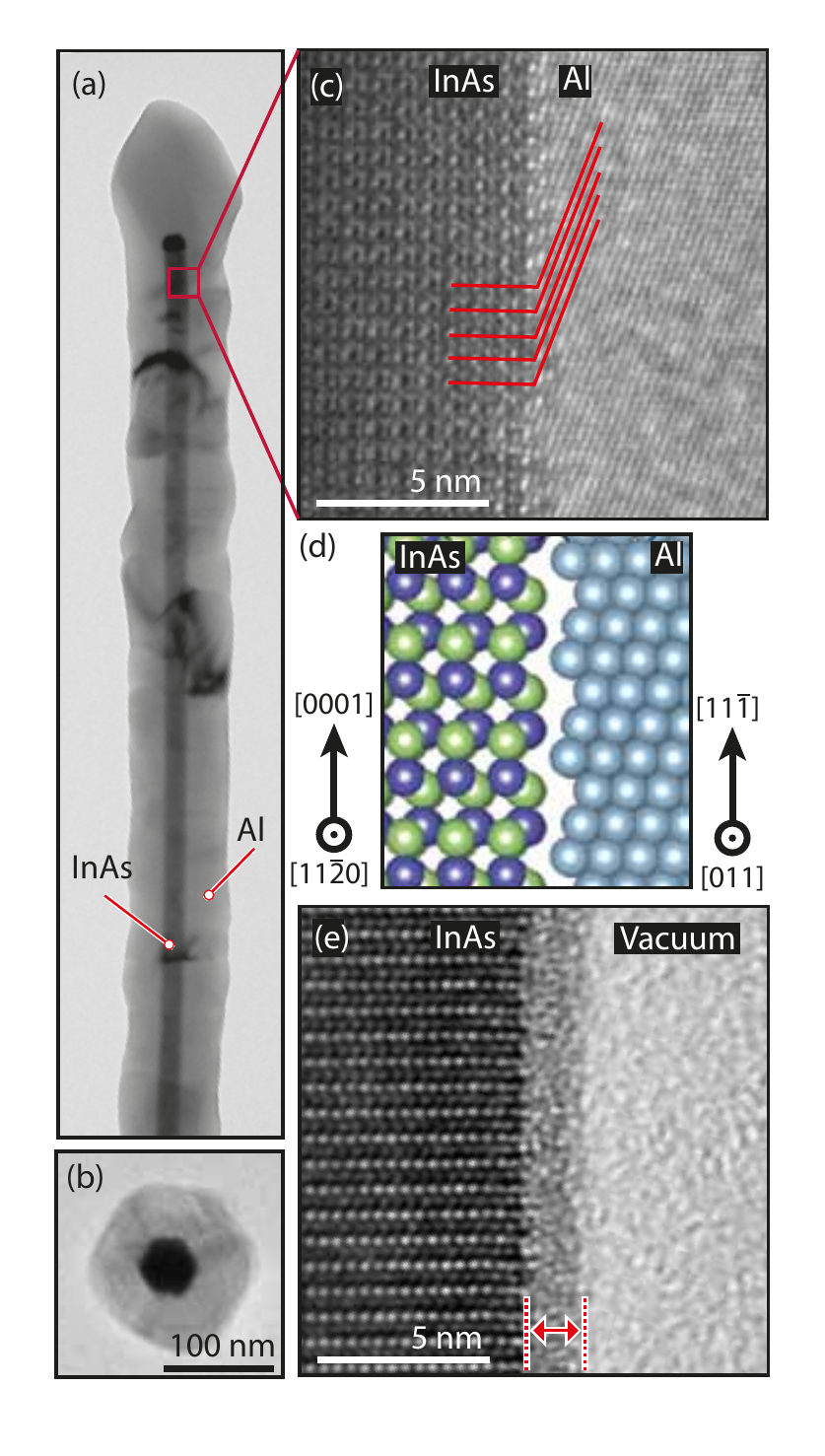}
        \caption{(a) Transmission electron micrograph of an InAs nanowire encapsulated in
MBE grown aluminum. (b) Top view TEM micrograph of a microtomed $\sim 100 \, \mathrm{nm}$ thick NW slice. The hexagonal cross-section of the wire is clearly seen and the aluminum shell conforms to this shape. (c) High-resolution TEM of the epitaxial interface between the InAs NW and the aluminum. No oxide is observed in the interface and the red lines indicate the alignment of the crystal planes. (d) Schematic illustration of the lattice matching of the InAs and the aluminum coating in the direction along the wire. (e) As (c) for a non-coated NW, showing the amorphous oxide on the surface (arrow).}
        \label{FIG:Fig1}
\end{figure}
Common for any such application of nanowires is the need for reproducible ohmic contacts, and this often becomes a pivotal step in the fabrication scheme. In this respect, the surface accumulation layer of InAs prevents the formation of a strong Shottky barrier at the metal/NW interface, making contacting easier than for most semiconductors. Nevertheless, to achieve contact, the oxide that covers the wire surface needs to be removed prior to metal deposition, and various techniques have been developed to tackle this problem. In the earliest works, the oxide was removed by a brief etch in hydrofluoric acid (HF) immediately prior to metal deposition\cite{Samuelson:2004}. This process is difficult to reproduce and often entire device batches turn out highly resistive due to oxide regrowth. To circumvent this problem Suyatin developed a procedure based on $\mathrm{(NH}_4\mathrm{)}_2\mathrm S_x$ to dissolve the oxide and passivate the surface thus protecting it against further oxidation\cite{Suyatin:2007}. This procedure has been very successful and is now widely adapted. Lately, also argon ion milling has been used with success\cite{Sourribes:2013}.\\
These approaches successfully removes the oxide, but also potentially etch or damage the InAs nanowire surface. Such microscopic interface degradation has so-far remained irrelevant as the techniques produce highly transparent ohmic contacts. Lately, however, due to a potential use in topological quantum information\cite{Alicea:2011}, there has been an intense interest in inducing superconductivity in strong spin-orbit coupling nanowires by virtue of the proximity effect. A recent theoretical study \cite{Takei:2013} argued that in order to achieve an induced superconducting gap free from sub-gap states (a \emph{hard} gap), a key parameter is the quality and uniformity of the SC/NW interface. Given the softness of the proximity gaps reported in the search for topologically protected Majorana quasi-particles so far\cite{Mourik:2012, Das:2012, Deng:2012, Churchill:2013}, alternative contacting schemes need to be investigated.\\
In his Letter we report a new approach for making electrical devices from InAs nanowires with oxide-free nanowire/metal epitaxial interfaces leading to reproducible low-resistance Ohmic electrical contacts. The method is based on a single molecular beam epitaxy (MBE) growth of an InAs nanowire core and a metallic aluminum (Al) nanowire shell. Because of the ultra-high vacuum of the MBE reactor the Al/InAs interface is highly uniform along the length of the wire and remains oxide-free leading to good electrical contact between the metal shell and the NW core. We demonstrate the subsequent fabrication of devices in a field-effect-transistor geometry by contacting the metallic aluminum shell and locally exposing the InAs channel, and characterize their electrical performance. We discuss the potential of this method for creating highly transparent superconducting contacts to nanowires and present electrical transport measurements below the critical temperature of the aluminum shell. We further demonstrate MBE-grown InAs nanowires coated by a partial shell thereby leaving half the nanowire susceptible to electrostatic gating while preserving the properties of the superconducting contact interface.\\
The InAs nanowires were grown on (111)B InAs substrates by the vapor-liquid-solid method in a solid-source MBE system\cite{Krogstrup:2009,Shtrikman:2009,Madsen:2013}. The wires grow in the [0001]B wurtzite crystal direction, perpendicular to the substrate. Following wire growth, without breaking vacuum ($10^{-11}$ Torr) in the growth chamber, the substrate is cooled below $-15 ^\circ \mathrm C$ and the growth is terminated with an aluminum layer of $50$-$100\,\mathrm{nm}$. At such low temperatures, the surface diffusion length of aluminum is only a few nanometers and the aluminum crystals are formed uniformly along the side facets of the InAs nanowire. Two types of InAs/Al core-shell structures were investigated, one where the substrate was rotated during the aluminum growth resulting in a complete shell, and one where the substrate rotation was disabled, resulting in wires with a half shell. In the following we first discuss the full-shell structures.\\
Figure 1(a) shows a transmission electron microscope (TEM) image of the resulting wires clearly revealing the InAs core and the surrounding aluminum shell. The crystal grains of the aluminum shell have an extension of $\sim $50-70$ \, \mathrm{nm}$ and are visible through the modulation of the TEM diffraction contrast resulting from different crystal-grain orientations. Due to the short diffusion length of the aluminum at low temperatures, the aluminum shell conforms to the hexagonal cross section of the InAs core. This is evident in Fig.\ 1(b), which shows a TEM micrograph of a thin cross section slice cut parallel to the growth substrate using an ultra microtome\cite{Wagner:2010}. Figure 1(c) shows a high-resolution TEM image of the InAs/Al interface from a different wire. The nanowire was oriented to have the [11$\bar 2$0]-direction along the viewing direction and InAs columns are clearly visible despite the aluminum shell. Importantly, the interface between the InAs core and the shell appears atomically abrupt with no intermediate oxide layer. This should be contrasted to the $2$-$3\, \mathrm{nm}$ oxide typically forming when an InAs nanowire is exposed to air as shown in Fig.\ \ref{FIG:Fig1}(e). Interestingly, crystal planes of the aluminum are also visible (red lines Fig.\ \ref{FIG:Fig1}(c)). As inferred from the corresponding fast Fourier transform (FFT, not shown) the Al is oriented with the [011]-direction along the viewing direction and [11$\bar 1$] along the [0001] NW axis. The visible ($\bar 1$1$\bar 1$)-planes of the aluminum (red lines) form an angle of 19.48$^\circ$ to the (1$\bar 1$00)/($\bar 2$1$\bar 1$) InAs/Al interface. The spacings of bulk ($\bar 1$1$\bar 1$) aluminum planes along the NW growth direction is thus $0.7015 \, \mathrm{nm}$, very close to the bulk InAs lattice spacing of $0.6995 \, \mathrm{nm}$. This gives a lattice mismatch of 0.3$\%$. This correspondence is indicated by the red lines in Fig.\ \ref{FIG:Fig1}(c) and schematically illustrated in Fig.\ \ref{FIG:Fig1}(d). Owing to the weak TEM contrast of aluminum and the local bending of the nanowire on the TEM support, it is not possible to resolve the aluminum crystal planes for every position along the nanowire. However, whenever the crystal planes could be resolved the above relationship to the InAs was observed. We note, that from bulk considerations, a similar lattice matching is not present in the direction perpendicular to the wire axis. The observed preferred orientation of the aluminum may, however, be a result of the crystal more easily accommodating strains along the narrow $\sim 35\, \mathrm{nm}$ facets than along the wire axis.
Most importantly, amorphous layers were never observed in the interface and the atomically sharp transition continues uniformly along the wire.\\
\begin{figure}[t]
        \centering
        \includegraphics[width=8cm]{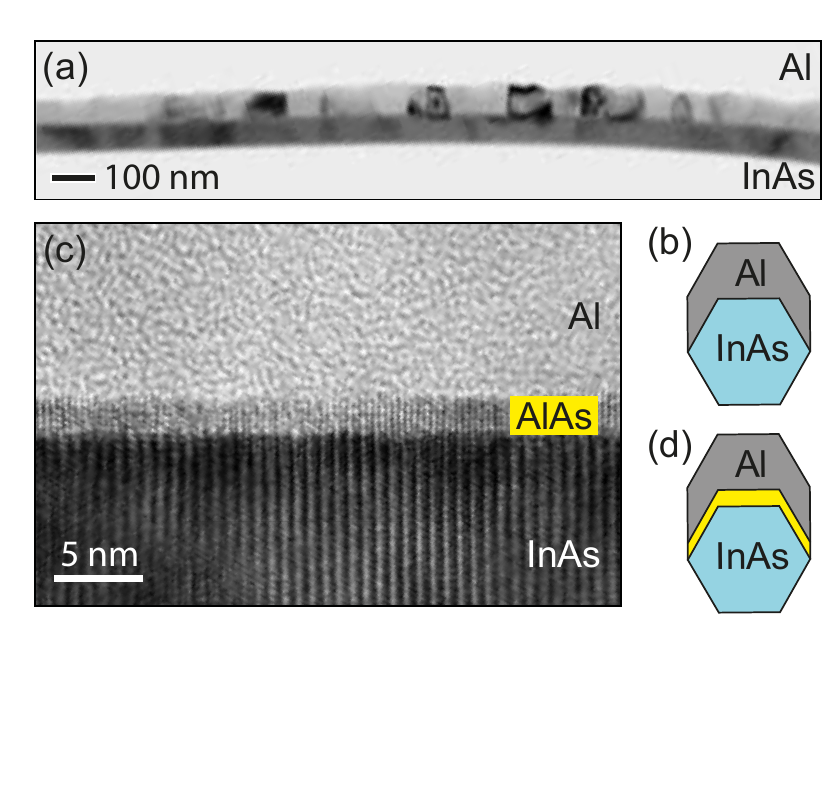}
        \caption{(a) TEM micrograph of an InAs nanowire coated on one side only by aluminum as schematically shown in panel (b). The Al/NW interface is identical to those presented in Fig.\ 1. (c) TEM micrograph of the Al/NW interface showing an example where a high band-gap AlAs tunnel barrier is grown to separate the nanowire from the contact as shown schematically in (d). Such structures have the potential for enhancing the tunability of the chemical potential in the wire and for optimizing the induced superconductivity.}
        \label{FIG:Fig4}
\end{figure}
The second type of wires we have grown uses the same growth technique as described above and takes advantage of the short surface diffusion length of aluminum at low temperatures to produce InAs NWs half covered with aluminum. This can be achieved by disabling the substrate rotation while depositing the aluminum. Further, because all wires are in epitaxial relation to the InAs growth-substrate the NWs can be oriented to all have the desired side-facets facing the aluminum source and typical results are shown in Fig.\ \ref{FIG:Fig4}(a,b). These half-covering aluminum contacts have the same interface properties as shown in Fig.\ \ref{FIG:Fig1} for the full shells.\\
While these results show that contacts produced inside the MBE reactor maintain uniform oxide-free NW/Al interfaces, our approach also allows the control of the MBE technique to be employed for engineering the details of the contact interfaces. Figures \ref{FIG:Fig4}(c,d) show an example where a $3\, \mathrm{nm}$ tunnel barrier of high band-gap AlAs was incorporated to separate the InAs core from the aluminum half-shell.\\
\begin{figure}
        \centering
        \includegraphics[width=8cm]{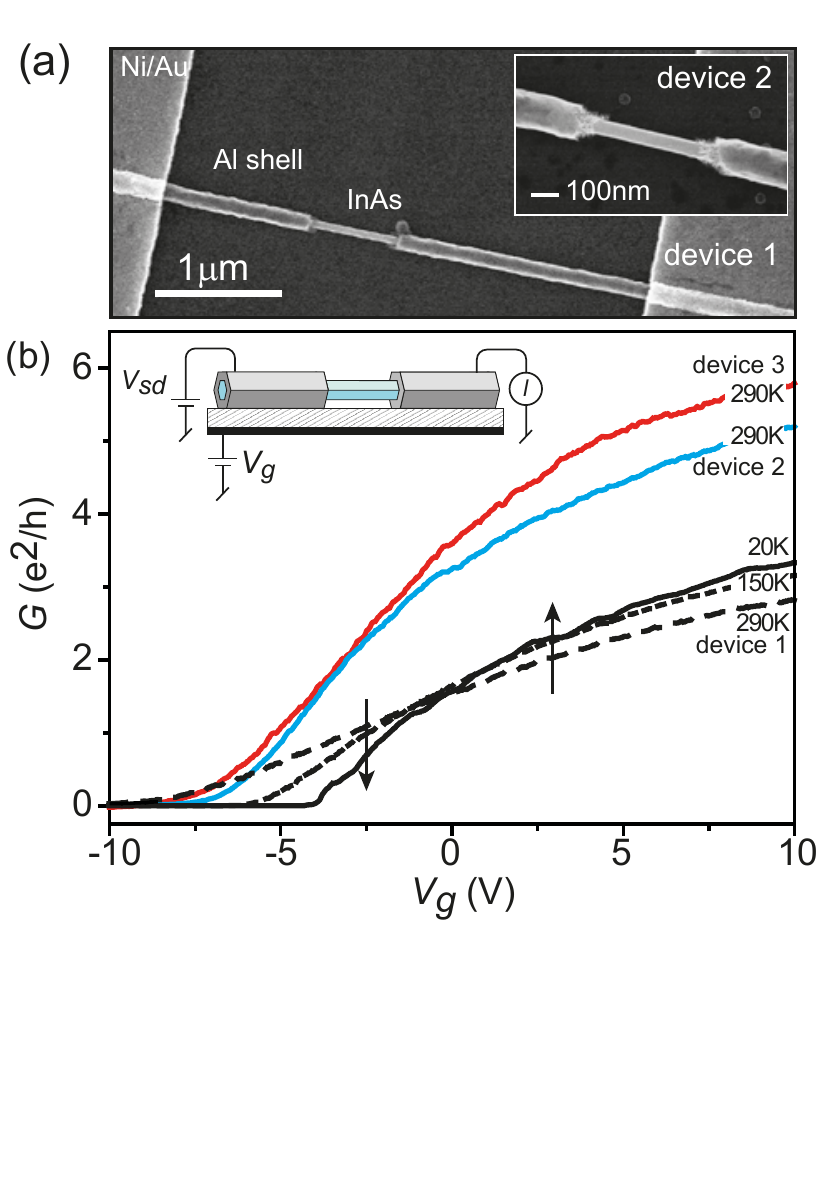}
        \caption{(a) Scanning electron micrograph of Device. The aluminum shell has been etched along a 600 nm segment to expose the InAs core. After the etch, the device is capped in 30 nm hafnium oxide and the doped substrate acts as a back gate electrode for modulation the carrier density in the InAs (inset to panel (b)). The inset shows the exposed channel of another device (device 2). (b) Conductance as a function of $V_g$ for three different devices. The temperature dependence is shown for device 1, showing an increase in conductance upon cooling down for $V_g > 0 \, \mathrm V$ (rightmost arrow).}
        \label{FIG:Fig2}
\end{figure}
\begin{figure}[t]
        \centering
        \includegraphics[width=8cm]{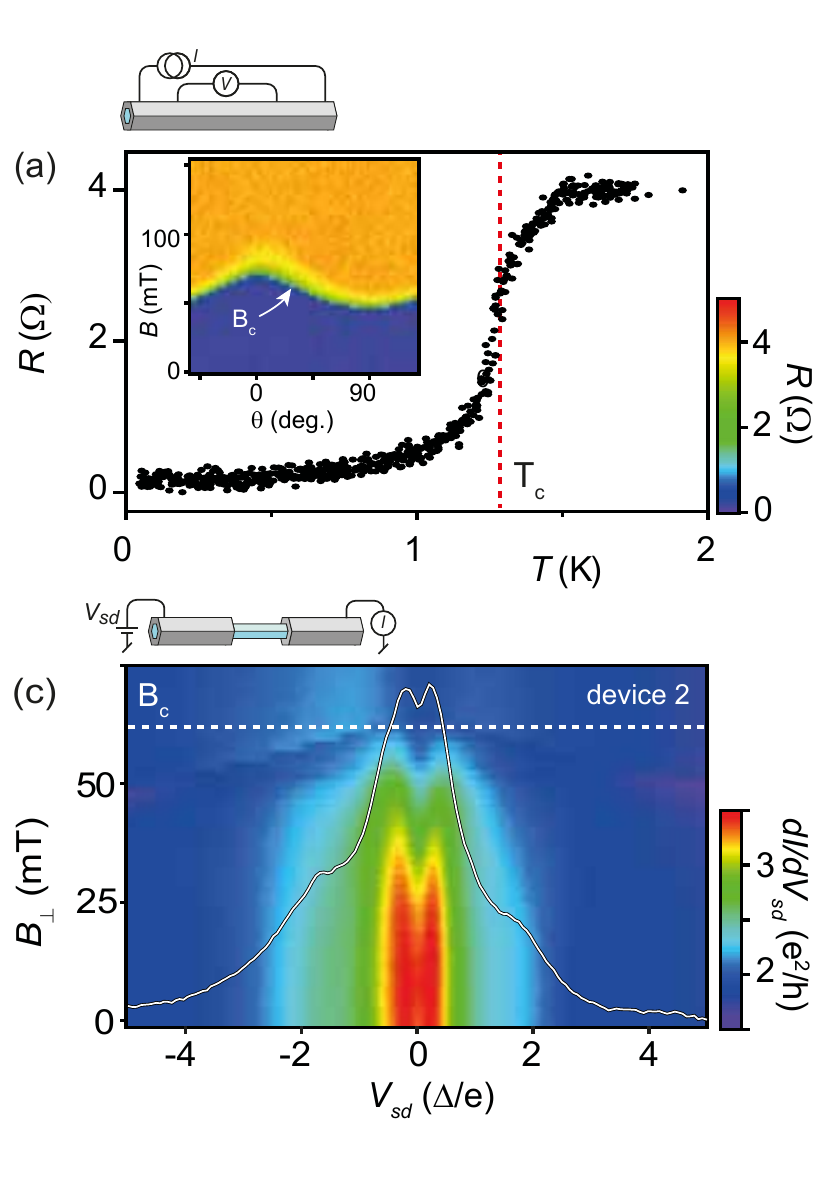}
        \caption{(a) Temperature dependence of the four-terminal resistance of a wire with an intact aluminum shell showing the superconducting transition at $T_c = 1.3 \, \mathrm{K}$. Inset shows corresponding four-terminal resistance at $T=20 \, \mathrm{mK}$ as a function of magnetic field strength and angle with respect to the wire axis. The critical field $B_c$ is indicated and is largest for the parallel alignment of field and axis. (b) Measurement of $dI/dV_{sd}$ vs.\ $V_{sd}$ and $B_\bot$ at $T=50 \, \mathrm{mK}$ for device 2 with a $\sim 600 \, \mathrm{nm}$ wide segment of exposed InAs (cf.\ Fig.\ 2). A clear conductance increase is observed for $|V_{sd}| \lesssim \pm 2\Delta/e = \pm 0.4 \, \mathrm{meV}$. White line: $B=0\,\mathrm T$ trace ($dI/dV_{sd} = 1.8(3.3)e^2/h$ for $V_{sd} = -1(0)\,\mathrm{mV}$).}
        \label{FIG:Fig3}
\end{figure}
We now focus on the electrical characterization of the contacts. To fabricate gateable electrical devices, the wires were liberated from the substrate by a brief sonication in methanol, and a small amount of the resulting suspension was deposited on a substrate of silicon capped with 500$\, \mathrm{nm}$ SiO$_2$. Wires were located with respect to predefined alignment marks using optical dark field microscopy, and the ends of the wires contacted using electron beam lithography and evaporation of Ni or Ti. To break through the native oxide present on the aluminum surface a brief argon ion milling was performed inside the evaporation chamber. To gain access to the InAs core, a second electron beam lithography step was performed to open narrow windows in a poly(methyl-methacrylate) resist between the contacts and the shell was removed by a 2-3 sec.\ etch in 12\% buffered HF\cite{Ecthantfodnote}. Finally, the device was coated in $20$-$30\, \mathrm{nm}$ of hafnium oxide using atomic layer deposition. Figure \ref{FIG:Fig2}(a) shows scanning electron micrographs of typical devices prior to the HfO$_2$ deposition, with a $\sim 0.5 \, \mu \mathrm{m}$ wide segment of the InAs core exposed. The inset to Fig.\ \ref{FIG:Fig2}(b) shows the device schematics. We note that in order to employ the aluminum shell to aid electrical contacts, the second lithography step is in principle obsolete as the first metal layer can act as the etch mask for exposing the InAs core; here, however, the two-step approach was used to separate the lithographically-defined contacts from the region of exposed NW to ensure that close to this region the superconducting properties of the shell are intact.\\
Figure \ref{FIG:Fig2}(b) shows the measured conductance as a function of the voltage $V_{\mathrm{g}}$ applied to the conducting back plane of the substrate for three different devices. As expected for undoped InAs NWs the devices act as $n$-type semiconductors with an increased conductance for increased $V_g$. The devices are pinched-off at $V_g = -10 \, \mathrm V$ and the peak conductance for $V_g = 10 \, \mathrm V$ is $2.8$, $5.1$, and $5.8 \, \mathrm{e}^2/\mathrm{h}$ for devices 1, 2 and 3, respectively.\\
Focusing on the low temperature characteristics Fig.\ \ref{FIG:Fig3}(a) shows the four-terminal resistance as a function of temperature of an aluminum-coated NW without the core exposed, showing a broad superconducting transition centered around $T_c \sim 1.3 \, \mathrm{K}$. The inset to Fig.\ \ref{FIG:Fig3}(a) shows the corresponding dependence of the resistance on magnetic field measured at $50 \, \mathrm{mK}$ as a function of angle $\theta$ of the field with respect to the wire axis. For this device the thickness of the aluminum is $\sim 50$-$70 \, \mathrm{nm}$ and the critical field peaks at $B_c^\| \approx 75 \, \mathrm{mT}$ for a field aligned along the nanowire axis and has a minimum value of $B_c^\bot \approx 50 \, \mathrm{mT}$ in the perpendicular orientation.\\
Figure \ref{FIG:Fig3}(b) shows a plot of $dI/dV_{sd}$ vs.\ $V_{sd}$ and $B_\bot$ for device 2 with the InAs core exposed along a $\sim 600 \, \mathrm{nm}$ segment. As a consequence of the superconducting contacts, a clear conductance increase is observed at low fields for $|V_{sd}| < 2\Delta/e$ disappearing when $B_\bot > B_c\sim 60 \, \mathrm{mT}$. \\
Turning to interpretation, we first note that the maximum conductance values obtained in Fig.\ \ref{FIG:Fig2} at $V_g= 10\, \mathrm V$ are comparable to the best results we have achieved for devices of comparable lengths and diameters using either HF etching,  $\mathrm{(NH}_4\mathrm{)}_2\mathrm S_x$ passivation, or argon milling for removing the InAs oxide. This indicates a barrier-free metal/semiconductor contact as is further supported by the temperature dependence of the $G$ vs.\ $V_g$ measurements shown in Fig.\ \ref{FIG:Fig2}(b), where the conductance increases upon cooling of the device (likely due to the reduction of phonon scattering) rather than decreasing as is most often observed for imperfect contacts due to the reduction of thermally excited transport over contact barriers. These results  confirm the expectations from the TEM analysis that the epitaxial interfaces result in a highly transparent electrical contacts being the first aim of our study.\\
Considering now the behavior at low temperature the results of Fig.\ \ref{FIG:Fig3}(a,b) show that the MBE deposited aluminum film behaves as expected for a mesoscopic superconductor. The transition temperature $T_c = 1.3 \, \mathrm{K}$ observed in Fig.\ \ref{FIG:Fig3}(a) is slightly higher than the bulk value for aluminum (1.2 K) and while the underlying mechanism is unknown, this behavior is a well established experimental fact for aluminum nanowires below typical dimensions of $\sim 50 \, \mathrm{nm}$\cite{Arutyunov:2008}. Also the width of the transition is typical for such superconductors, caused by sample inhomogeneity or phase slips\cite{Arutyunov:2008, Zgirski:2007}.\\
The critical magnetic field is related to the effective projected thickness of the aluminum shell to the direction of the field: $50$-$70\,\mathrm{nm}$ and $180$-$220\,\mathrm{nm}$ for the parallel and perpendicular fields. Measurements of $B_c$ for thin planar aluminum films\cite{Meservey:1971} yield $\sim 150$-$250 \, \mathrm{mT}$ and $\sim30$-$60 \, \mathrm{mT}$ for similar thicknesses. While the $B_c^\bot$ falls within this expected range, our measured $B_c^\|$ is lower. The reason may be related to thickness variations due to the finite grain size of the aluminum (Fig.\ \ref{FIG:Fig1}(a)) which has a relatively larger effect on $B_c^\|$ than on $B_c^\bot$. On other nanowire samples with a shell thickness of $13 \, \mathrm{nm}$ we have measured $B_c^\| \sim 1900 \, \mathrm{mT}$ and $B_c^\bot \sim 150 \, \mathrm{mT}$ matching the expectations from thin planar films\cite{Elsewhere}.\\
The data of Fig.\ \ref{FIG:Fig3}(b) for device 2 confirms that the epitaxial aluminum can act as superconducting contacts. The superconductivity appears as an increased conductance for $|V_{sd}| \lesssim 2\Delta/e$ where $\Delta$ is the gap energy for bulk aluminum. The absence of a zero-bias supercurrent can likely be attributed to inadequate filtering of the electrical wiring in the cryostat. The finite bias peaks and shoulders observed in the line-trace at $B=0 \, \mathrm{mT}$ is attributed to effects of Andreev reflections.\\
As mentioned above, the softness of the proximity induced gaps in nanowires contacted to superconductors in the traditional ways, is a major problem in the ongoing experimental efforts to realize and study Majorana fermions in nanowires. To measure the characteristics of the gap in the present structures, hybrid normal-metal/NW/S devices must be realized in the tunneling regime and such devices will be the focus of future experiments. The TEM analysis, however, shows that the interface homogeneity is optimal, being the most important parameter to ensure a hard gap as suggested by recent theory\cite{Takei:2013}. Further, in realizing Majorana devices, tuning of the chemical potential in the contacted region is required and therefore the wire must be susceptible to electrostatic gating. Due to electrostatic screening from the contact a fully covering shell is therefore unsuited. In Ref.\ \cite{Mourik:2012} gatability was enabled by fabricating half-covering superconductors using advanced lithography aligned along the length of the wire. With the half-shell wires presented in Fig.\ \ref{FIG:Fig4} we have shown that with our approach, gatable structures with perfect interfaces can be created without the need of post processing. However, even with a part of the wire exposed to electrostatic gating, it remains unknown to which degree the chemical potential can be electrostatically tuned in a nanowire in direct contact to a superconductor, and it may be advantageous to enhance the gatablity by separating the two by a thin, uniform tunnel barrier as demonstrated in Fig.\ \ref{FIG:Fig4}. Moreover, it has been theoretically suggested\cite{Stanescu:2013} that a finite tunnel barrier may enhance the magnitude of the induced proximity gap for a nanowires with diameters beyond $\approx 40 \, \mathrm{nm}$. Finally, using thin aluminum shells to increase $B_c$ to $1.9 \, \mathrm{T}$ clearly enables the study of the topologically non-trivial regime in InAs which requires fields exceeding $\sim 400 \, \mathrm{mT}$, assuming a typical value of 10 for the InAs $g$-factor\cite{Bjork:2005, Jespersen:2006} and an induced gap equal to the gap of bulk aluminum.\\
In conclusion we have developed a scheme for producing controllable metal contacts to MBE-grown semiconducting nanowires having perfect oxide-free interfaces. The methods for subsequent fabrication of devices have been developed and the contact characteristics was studied by transmission electron microscopy and by low temperature electrical measurements. The method is generally applicable to all situations where electrical contacts to semiconductor nanowires are needed and transforms the contacting problem into the task of making metal/metal contact. The method is particularly promising for situations where the homogeneity of the metal-semiconductor interface is important such as when transparent, uniform coupling to superconductors is required, as expected, for instance, in creating long lived Majorana states. Half covering contacts are also demonstrated exposing the InAs core to electrostatic gating and we have found that with a thin $13 \, \mathrm{nm}$  aluminum shell the critical magnetic field is increased to $1.9$ Tesla. Whether these structures solves the problem of soft, small proximity induced superconducting gaps in the nanowires will be the focus of future experiments.\\
We acknowledge B. Wenzell and L. Schulte for TEM sample preparation and financial support by EU FP7 project SE2ND (no. 271554), the Danish Strategic Research Council, the Danish Advanced Technology Foundation, the Carlsberg Foundation, the Lundbeck Foundation, and Microsoft Project Q. The Center for Quantum Devices is supported by the Danish National Research Foundation. 
\bibliography{ZiinoBIB}

\begin{mcitethebibliography}{29}
\bibitem[Ect()]{Ecthantfodnote}
Other aluminum etchants can also be used, however, we find that that HF is
  superior for making sharp Al edges\relax
\mciteBstWouldAddEndPuncttrue
\mciteSetBstMidEndSepPunct{\mcitedefaultmidpunct}
{\mcitedefaultendpunct}{\mcitedefaultseppunct}\relax
\EndOfBibitem
\bibitem[Els()]{Elsewhere}
Once the critical field is increased to allow half a flux quantum to thread the
  superconducting cylinder the full-shell devices exhibit pronounced
  Little-Parks oscillations. The results will be presented elsewhere.\relax
\mciteBstWouldAddEndPunctfalse
\mciteSetBstMidEndSepPunct{\mcitedefaultmidpunct}
{}{\mcitedefaultseppunct}\relax
\EndOfBibitem
\end{mcitethebibliography}
\end{document}